\documentclass[twocolumn,showpacs,preprintnumbers,amsmath,amssymb,prb]{revtex4}
\usepackage{graphicx}
\usepackage{dcolumn}
\usepackage{color}
\usepackage{bm}
\usepackage{float}
\usepackage{gensymb}
\begin{document}

\title{Ab initio study of structural stability of small 3$d$ late transition metal clusters : Interplay of magnetization and hybridization}

\author{Soumendu Datta,$^1$ Mukul Kabir,$^2$ and Tanusri Saha-Dasgupta$^1$}

\affiliation{$^1$ Advanced Materials Research Unit and Department of Material Sciences, S.N. Bose National Centre for Basic Sciences, JD Block, Sector-III, Salt Lake
City, Kolkata 700 098, India \\
$^2$Department of Materials Science and Engineering, Massachusetts Institute of Technology, Cambridge, Massachusetts 02139, USA }

\date{\today}

\begin{abstract}
Using first-principles density functional theory based calculations, we analyze the structural stability of small clusters of 3$d$ late transition metals. We consider the relative stability of the two structures - layer-like structure with hexagonal closed packed stacking and more compact structure of icosahedral symmetry. We find that the Co clusters show an unusual stability in hexagonal symmetry compared to the small clusters of other members which are found to stabilize in icosahedral symmetry based structure. Our study reveals that this is driven by the interplay between the magnetic energy gain and the gain in covalency through $s$-$d$ hybridization effect. Although we have focused our study primarily on clusters of size 19 atoms, we find that this behavior to be general for clusters having sizes between 15 and 20.
\end{abstract}
\maketitle

\section{\label{sec:intro}Introduction}
Atomic clusters of nanometer size have attracted a special attention of the present day research due to their applications in the field of optoelectronic,\cite{ref1,ref2} catalysis,\cite{ref3,ref4} data-storage,\cite{ref5} sensors\cite{ref6,ref7,ref8,ref9} etc. The first step to the theoretical study of the properties of clusters is the determination of the minimum energy structures. The equilibrium minimum energy structures of small clusters often prefer compact geometries\cite{ref10} like icosahedral or cub-octahedral symmetry based structures. It has also been found in several cases that the deformed three dimensional sections of the face centered cubic (fcc) or hexagonal closed packed (hcp) lattice appear as degenerate energy state or closely lying isomers.\cite{ref11} However, depending upon the local symmetry, they exhibit profoundly different properties. For example, planner gold clusters exhibit outstanding catalytic activity compared to their bulk counterpart of fcc symmetry,\cite{ref12} bi-layer Ru-nanoclusters exhibit significant chemical activity towards H$_2$O splitting compared to Ru-clusters of hcp symmetry.\cite{ref13}  Similarly, the dependence of magnetic behavior of the Pd clusters on cluster symmetry is found to be significant.\cite{ref15} All of these indicate that the determination of the local symmetry is an unavoidable part in a cluster calculation.

  In this article, we have performed a first principles based analysis to understand the structural trend of transition metal clusters. Transition metal clusters demand special attention because of their fascinating magnetic properties,\cite{new1,new2} the dependence of the equilibrium structure on magnetism\cite{new3} as well as their potential biomedical applications.\cite{new4,new5,new6} We focus our attention only to the 3$d$ late transition metal clusters. Among the 3$d$ late transition metal elements, Mn has a half-filled $d$-level, while the others have more than half-filled $d$-level. Considering the earlier studies on structure of the 3$d$ late transition metal clusters, it is seen that the small Mn and Fe clusters generally prefer a compact icosahedral growth pattern as has been shown by the first principles calculations for the Mn clusters\cite{ref16,ref17,ref18,ref19,ref20} and Fe clusters.\cite{ref21,ref22,ref23,ref24} For relatively less magnetic Ni clusters and nonmagnetic Cu-clusters, the first principles calculations and also some experimental evidences indicate mainly icosahedral growth pattern.\cite{ref25,ref26,ref27,ref28,ref29}$^,$\cite{ref30,ref31,ref32,ref33,ref34,ref35} Some recent calculations\cite{ref36,ref37,ref38} also highlight non-icosahedral or amorphous structural pattern for small clusters of coinage metals like Cu, Ag and Au. On the other hand, the ferromagnetic small Cobalt cluster is quite different from the other members of the 3$d$ late transition metal series, particularly Mn and Fe clusters. The small Co$_n$ clusters rather prefer relatively non-compact layer-like structures. In our recent work\cite{ref39} using first-principles density functional study, we have shown a clear hexagonal growth pattern for the small Co$_n$ clusters (15$\leq$n$\leq$20). Hexagonal symmetry based structures in this size range consist of 3-planes with hcp stacking. Also recently, this layer-like structures of the small Co$_n$ clusters (13$\leq$n$\leq$23) has been reported by Gong {\it et al}\cite{new7,ref40} using density functional calculation. Experimental work on small Co$_n$ clusters (n$<$50),\cite{ref41,ref42,ref43} though is unable to give any definitive conclusion, indicates non-icosahedral packing too.

 It is therefore curious why the small cobalt clusters prefer a hcp growth pattern with layer-like stacking, while the clusters of the other 3$d$ late transition metal elements apparently prefer a more compact icosahedral growth pattern. In order to have an understanding on this issue, we have chosen these two close packed structures with hcp and icosahedral symmetries as the starting guesses and allowed them to relax under the assumption of {\it collinear} magnetic ordering. We have studied the relative stability between these two symmetry based structures in terms of energetics, structural and electronic properties. We have carried out our study for the entire series of 3$d$ late transition metal clusters i.e  Mn$_n$, Fe$_n$, Co$_n$, Ni$_n$ and Cu$_n$. Our study reveals that the contrasting behavior of the stability of Co-clusters compared to the other members, arises due to the interesting interplay of the effects of magnetization and hybridization.

\section{\label{sec:methodology} Computational Details}
We employed density functional theory within plane wave pseudo potential method as implemented in Vienna {\it ab initio} simulation package.\cite{ref44} We used the projected augmented wave pseudo-potentials\cite{ref45,ref46} and the Perdew-Bruke-Ernzerhof exchange-correlation functional\cite{ref47} of the generalized gradient approximation (GGA). The pseudo potentials for the transition metal elements studied in this work, were generated considering the 3$d$ and 4$s$ electrons as the valence electrons. The energy cut-off was 335 eV for the cluster calculation of each transition metal. We did both spin-polarized and non-spin polarized calculations at the $\Gamma$ point of the Brillouin zone. Geometry optimizations have been performed using the conjugate gradient and the quasi-Newtonian methods until all the force components were less than a threshold value of 0.01 eV/{\AA}. For the cluster calculation, a simple cubic super-cell was used with periodic boundary conditions, where two neighboring clusters were kept separated by around 12 {\AA} vacuum space, which essentially makes the interaction between cluster images negligible. To determine the magnetic moment of the minimum energy structure in spin-polarized calculations, we have explicitly considered all the possible spin multiplicities for each structure under the approximation of {\it collinear} atomic spin arrangements.

\section{Results and Discussions}
\begin{figure}
\begin{center}
\rotatebox{0}{\includegraphics[height=5.0cm,keepaspectratio]{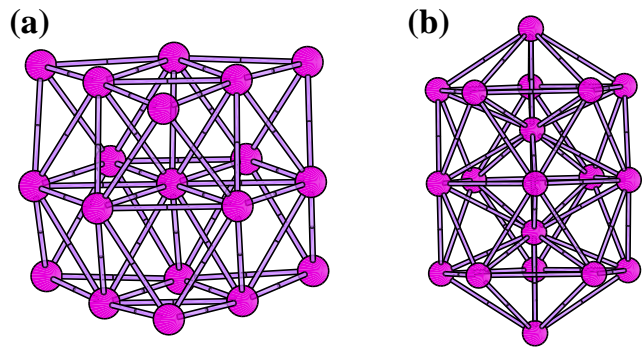}}
\caption{(Color online) Cluster structures of 19 atoms with (a) hexagonal closed packed and (b) icosahedral symmetries. These two competing structural symmetries have been
considered in this work to determine the minimum energy structure for each X$_{19}$ cluster, $X$=Mn, Fe, Co, Ni and Cu.}
\label{tseries_str}
\end{center}
\end{figure}

To make our conclusion regarding structural and electronic properties of the clusters more general, we have chosen the cluster size of 19 atoms for each 3$d$ late transition metal cluster considered in this study, instead of cluster size of 13-atoms, which is the first geometric magic size for the icosahedral symmetry based structure of the most transition metal clusters. Fig. \ref{tseries_str} shows the cluster structures of 19 atoms with hcp and icosahedral symmetries. A 19-atoms cluster structure with hcp symmetry, can be viewed as stacking of three planes containing 6, 7 and 6 atoms respectively in each of these planes. On the other hand, a 19-atoms cluster structure with icosahedral symmetry can be thought of two interpenetrating 13-atoms icosahedrons. As seen from Fig.\ref{tseries_str}, a 19-atoms cluster structure of hcp symmetry appears more open in the sense of lesser value of average co-ordination of atoms as well as more layer-like structure compared to the 19-atoms icosahedral structure. In the first step of our optimization procedure, we have started with ideal hexagonal and icosahedral structure for 19 atom clusters and have
optimized them. In the second step, we have randomly displaced few atoms in the optimized structure, obtained in the first step, and have
reoptimized to get the ``final optimized'' structure. The second step has been carried out considering all possible collinear spin arrangements
of the atomic spins in each X$_{19}$ (X = Mn, Fe, Co, Ni, Cu) cluster within the spin-polarized calculation. The optimized clusters do not
have perfect hexagonal or icosahedral symmetry but are heavily distorted.

\begin{figure}[b]
\begin{center}
\rotatebox{0}{\includegraphics[height=5.9cm,keepaspectratio]{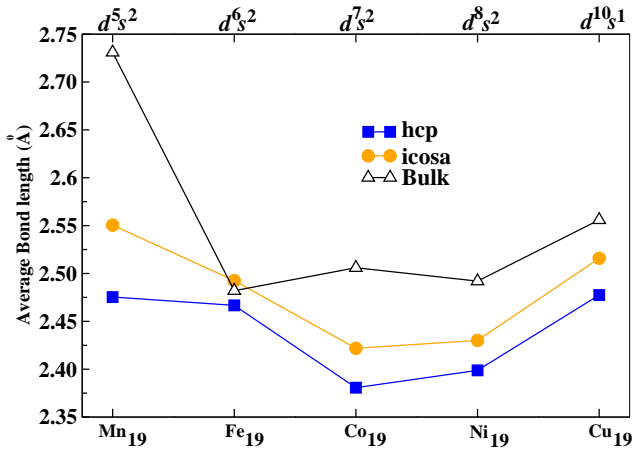}}
\caption{(Color online) Plot of the average nearest-neighbor bond lengths $\langle$r$\rangle$ (see text) for the optimized hcp and icosahedral 19-atoms clusters of Mn, Fe, Co, Ni and Cu. Blue (dark) squares correspond to the data points for the hcp structure and orange (light) circles for data points of the icosahedral structure of each X$_{19}$ cluster in spin-polarized calculation. The corresponding bulk-values have been shown with empty triangles. The atomic valance electronic configuration for each element has been marked at the top of the figure. }
\label{avgbl}
\end{center}
\end{figure}

\begin{figure}
\begin{center}
\rotatebox{0}{\includegraphics[height=5.9cm,keepaspectratio]{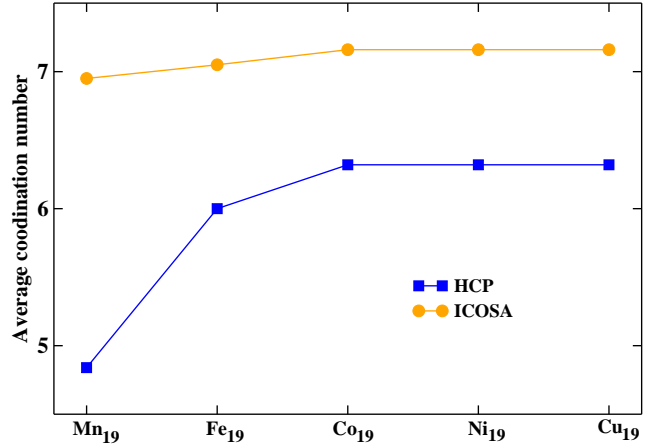}}
\caption{(Color online) Plot of the average co-ordination number for the optimized hcp and icosahedral 19-atoms clusters of Mn, Fe, Co, Ni and Cu.}
\label{coord}
\end{center}
\end{figure}

 First, to analyze the optimized structures, we define the average nearest-neighbor bond length as $\langle r \rangle$  = $\frac{1}{n_b}\sum_{i>j}r_{ij} $, where $r_{ij}$  is the bond distance between the $j$-th and $i$-th atoms, and $n_b$ is the number of such bonds. In cluster calculation, we considered that the two atoms are said to be bonded if their inter atomic distance is within 2.75 \r{A}, which is larger than any of the nearest neighbor bulk bond-lengths of these 3$d$ late transition metal elements. Fig. \ref{avgbl} shows the plot of the average nearest-neighbor bond-lengths of the optimized structures of X$_{19}$ clusters for both the symmetries. It is seen that the average nearest-neighbor bond-lengths for the hcp structures are consistently smaller than those of the icosahedral counterpart in agreement with the previous study,\cite{ref53} which indicates that the net attraction of nucleus on outer shell electrons is effectively more for the hcp symmetry based structure. As the $d$-shell gets filled one by one electron from Mn $\rightarrow$ Fe $\rightarrow$ Co $\rightarrow$ Ni $\rightarrow$ Cu, the ion-electron interaction gets stronger, which increases the binding. On the other hand, electron-electron repulsion also increases, which starts to downplay the gain in electron-ion attraction. On top of this effect, the increased atom-centered magnetic moments also play a significant role, specially for the members left of Co along Co$_{19}$ to Fe$_{19}$ to Mn$_{19}$.

\begin{table}[]

\begin{center}
\caption{Total binding energies and total magnetic moments of the minimum energy structure of hcp and icosahedral symmetries for each X$_{19}$ cluster, $X$=Mn, Fe, Co, Ni and Cu in spin-polarized calculations.}
\vskip 0.2cm
\begin{tabular}{ccccccc}
\hline
\hline
Clusters & & \multicolumn{2}{c} {Binding energy (eV)}& & \multicolumn{2}{c}{Magnetic moment ($\mu_B$)} \\
\cline{3-4}\cline{6-7}
           & &        hcp        &     icosa          & & hcp       &   icosa             \\
\hline
\\
Mn$_{19}$     & &     43.87        &     45.12         & &  15         &  19  \\
Fe$_{19}$     & &     64.35        &     66.26         & &  58        &  58  \\
Co$_{19}$     & &     72.01        &     70.80         & &  39        &  37  \\
Ni$_{19}$     & &     65.13        &     65.13         & &  18        &  14   \\
Cu$_{19}$     & &     47.04        &     46.95         & &  0         &  0  \\
\hline
\end{tabular}
\end{center}
\end{table}

 In Fig. 3, we further show the average co-ordination number plotted for the optimized structures of X$_{19}$ clusters in both symmetries. We find that average co-ordination for hcp symmetry based structure is less than that of icosahedral symmetry based structure, giving rise to a more open geometry although average bond length is smaller for hcp based structure compared to icosahedral structure (cf. Fig. 2).

\begin{figure*}[]
\begin{center}
\rotatebox{0}{\includegraphics[height=15.6cm,keepaspectratio]{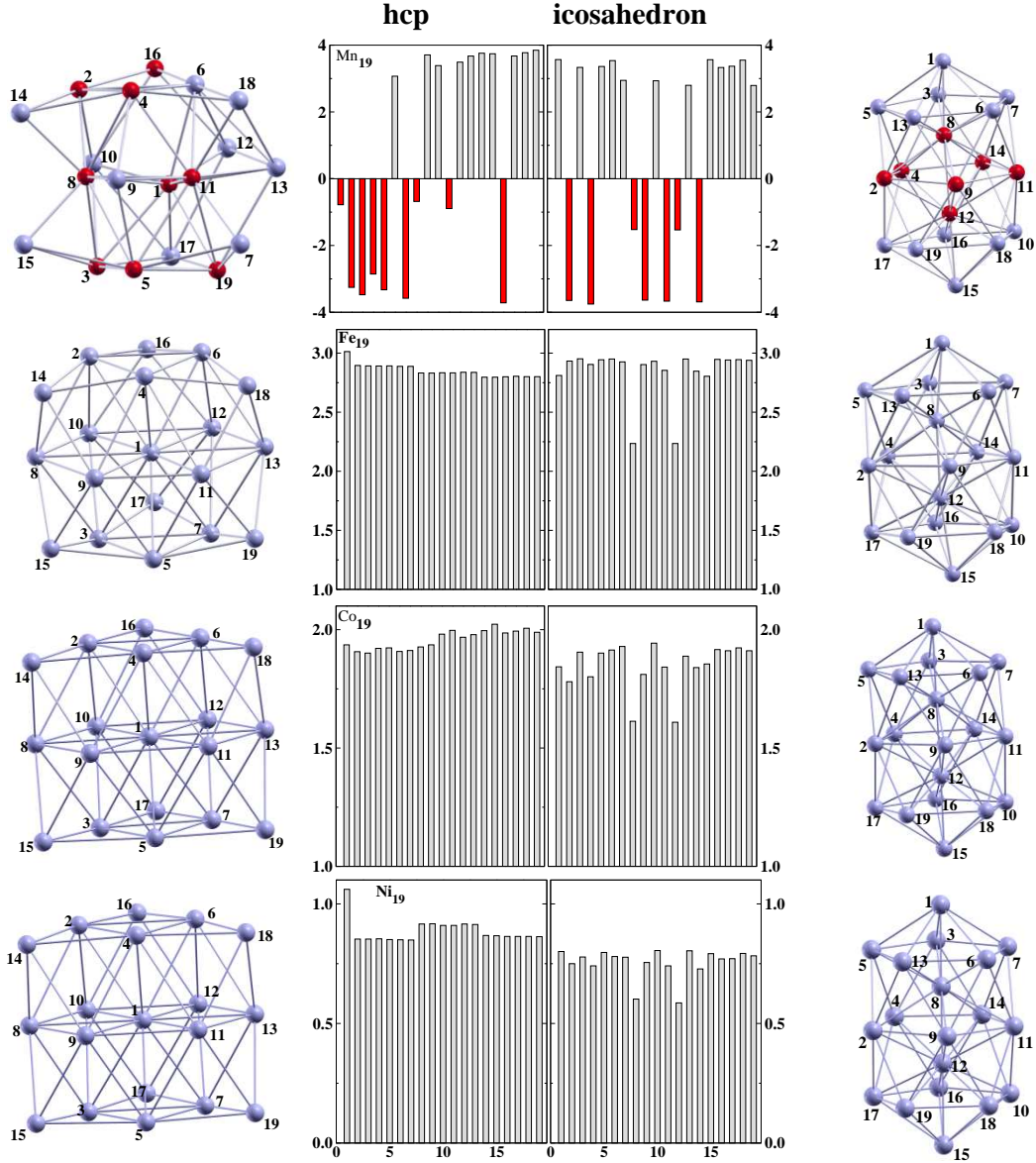}}
\caption{(Color online) Structures and atomic magnetic moment distribution in the optimized hcp and optimized icosahedral 
structures of Mn$_{19}$, Fe$_{19}$, Co$_{19}$ and Ni$_{19}$ clusters in spin polarized calculation. For Mn$_{19}$ clusters,  gray color 
represents up or positive and red (dark gray) color represents down or negative magnetic moment. For each of Fe$_{19}$, Co$_{19}$ and 
Ni$_{19}$ clusters, the atoms are ferromagnetically coupled, each with positive or up magnetic moments and are, therefore, represented 
by same color (gray).  The individual atomic magnetic moments of the constituent atoms in each optimized  structure, have been represented by bar plots, length of the bars corresponding  to the magnitude of atomic moments. The numbering of the atoms in each structure 
is indicated in the plots of the structures shown as insets. }
\label{magdis}
\end{center}
\end{figure*}

Binding energy for each X$_{19}$ cluster is calculated as $E_B(X_{19}) = [19E(X) - E(X_{19})]$, where E(X) and E(X$_{19}$) are the total energy of an isolated $X$ atom and that of a X$_{19}$ cluster respectively. In such a definition, a positive sign in $E_B$ corresponds to binding. Table I shows the total binding energy and the total magnetic moment of the optimized hcp and icosahedral structures of each X$_{19}$ cluster. It is seen that the icosahedral symmetry based structure is stabler than the hcp symmetry based structure for Mn$_{19}$ and Fe$_{19}$ clusters. Conversely, the hcp symmetry based structure is energetically more favorable than the icosahedral structure for the Co$_{19}$ cluster, while both the structures are almost degenerate for the Ni$_{19}$ and Cu$_{19}$ clusters within the accuracy of our calculations. Analyzing the atomic spin orientations in the optimized structures of both the symmetries of each X$_{19}$ cluster, we found that the Mn-Mn interactions within the Mn$_{19}$ cluster is mostly antiferomagnetic for both the optimal hcp and optimal icosahedral phases as mentioned by the earlier works.\cite{ref16,ref19} On the other hand, each of the Fe$_{19}$, Co$_{19}$ and Ni$_{19}$ clusters is ferromagnetic for either of the two structural symmetries, with decreasing total magnetic moment because the atom centered magnetic moments decrease as one goes along Fe$_{19}$ $\rightarrow$ Co$_{19}$ $\rightarrow$ Ni$_{19}$. Fig. \ref{magdis} shows the distribution of atomic magnetic moments of each of the X$_{19}$ clusters derived from both the symmetries together with the corresponding optimized structures. Note that bulk Mn is also anti-ferromagnetic and bulk Fe, Co, Ni are ferromagnetic (with magnetic moment per atom 2.2 $\mu_B$ for Fe,\cite{ref48,ref49} 1.72 $\mu_B$ for Co\cite{ref49} and 0.616 $\mu_B$ for Ni\cite{ref50}). The Cu$_{19}$ cluster is found to be non-magnetic with zero magnetic moment. 

\begin{figure}[]
\begin{center}
\rotatebox{0}{\includegraphics[height=4.8cm,keepaspectratio]{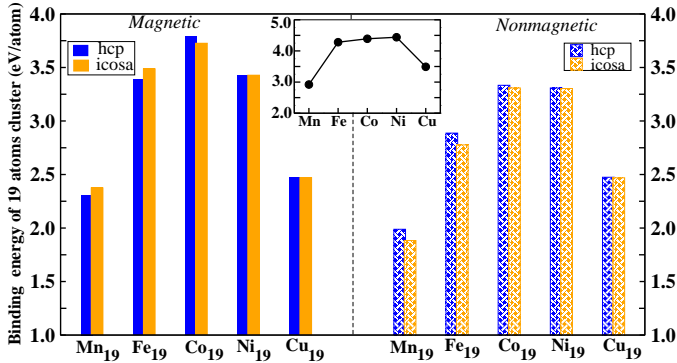}}
\caption{(Color online) Plot of binding energy per atom of the minimum energy hcp and icosahedral structures for each X$_{19}$ ($X$= Mn, Fe, Co, Ni, Cu) cluster in both spin polarized calculation (left) and non-spin polarized calculation (right). The blue (dark) vertical bars correspond to binding energies of the minimum energy hcp structure and vertical orange (light) bars correspond to the binding energies of the minimum energy icosahedral structure. The inset shows the trend in the bulk binding energy per atom for Mn, Fe, Co, Ni and Cu. }
\label{cohesive}
\end{center}
\end{figure}

To have a visual representation, we have plotted the binding energy per atom for the optimal hcp and optimal icosahedral structures of each X$_{19}$ cluster in Fig. \ref{cohesive}. To understand the effect of 
magnetism on stability, we have also performed the non-spin polarized calculation for each X$_{19}$ cluster. The binding energies for the optimal hcp and icosahedral structures of X$_{19}$ clusters in non-spin polarized calculation are also shown in Fig. \ref{cohesive} (right panel) with shaded bars. Interestingly, non-spin polarized calculation shows that the Mn$_{19}$, Fe$_{19}$ and Co$_{19}$ clusters all stabilize in hcp symmetry based structures, while both the structures are again degenerate for the Ni$_{19}$ and Cu$_{19}$ clusters. From Fig. \ref{cohesive}, it is also clearly seen that the magnetic phase always has the higher binding energy for both the structures of each X$_{19}$ cluster, indicating that the magnetic phase is the stable phase for both the structures. Only in case of the Cu$_{19}$ cluster, the binding energy of each structure is same for both the spin polarized as well as non-spin polarized calculations, indicating that the Cu$_{19}$ cluster is nonmagnetic. From the binding energy plot, we therefore conclude that the magnetism switches the stable phase from hcp to icosahedron in case of the Mn$_{19}$ and Fe$_{19}$ clusters, while the magnetism enhances further the stability of the hcp phase for the Co$_{19}$ cluster. For the Ni$_{19}$ and Cu$_{19}$ clusters, the effect of magnetism is small and both the hcp as well as icosahedral symmetry based structures are almost degenerate for both spin polarized as well as non-spin polarized calculations. 

For Mn clusters, the effect of noncollinearity has been discussed in literature.\cite{ref51} Mn is prone to noncollinearity due to the presence of competing nature
of magnetic interactions, though the degree of noncollinearity is found to decrease for cluster sizes larger than 13.\cite{ref51} For Fe
and Co clusters, the degree of noncollinearity is reported to be further smaller compared to Mn.\cite{noncolinear} Noncollinearity is favored by the magnetic energy
associated with larger magnetic moments, which competes with chemical bonding energy. One would therefore expect reduction of noncollinearity in moving to
larger cluster size as well as moving from Mn to Fe and Co. However, in order to check the influence of the possible noncollinearity which may arise due to
competing magnetic interactions as well as orbital component of magnetic moment, driven by spin-orbit (SO) coupling, we have repeated our calculations for Fe$_{19}$
and Co$_{19}$ clusters in terms of GGA+SO calculations. The obtained results 
indicate that the Fe$_{19}$ and Co$_{19}$ clusters are essentially collinear, with degree of noncollinearity being less than 2$^{o}$ in agreement with that 
reported previously.\cite{noncolinear} Though the orbital components of magnetism are found to be of finite values $\approx$ 0.08 $\mu_B$, importantly the 
calculations carried out considering noncollinearity leads to only small changes in the binding energy differences of the icosahedral and hexagonal geometries 
by 1- 2 $\%$, keeping the main conclusion of our study unchanged. In the following, we focus primarily on the Fe$_{19}$ and Co$_{19}$ clusters, for which the 
switching of the  stable phase between hcp and icosahedral structures occurs.

\begin{figure}
\begin{center}
\rotatebox{0}{\includegraphics[height=5.7cm,keepaspectratio]{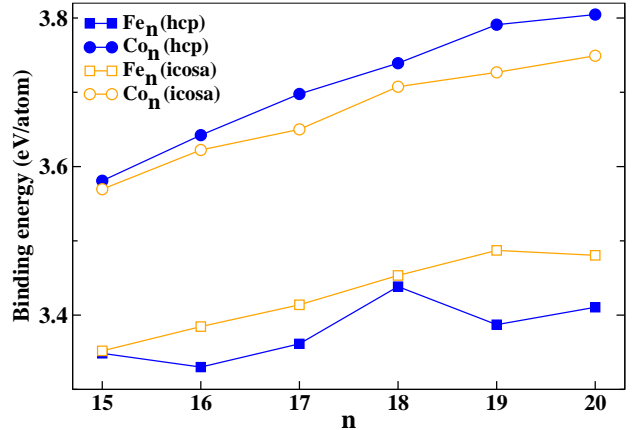}}
\caption{(Color online) Plot of binding energies of the minimum energy hcp and icosahedral structures for each Fe$_n$ and Co$_n$ cluster (15$\leq$n$\leq$20) in spin-polarized calculations. The squares correspond to the data points for the Fe$_n$ clusters (solid squares for the hcp symmetry and empty squares for the icosahedral symmetry) and the circles for the data points of Co$_n$ clusters (solid circles for the optimal hcp structure and empty circles for the optimal icosahedral symmetry).}
\label{13to20}
\end{center}
\end{figure}

 It is important to note that the trend in binding energy calculation is very robust, being independent of the type of pseudo-potential or the nature of the exchange-correlation functional used in this study. We also found that this trend to be general for clusters having sizes 15$\leq$n$\leq$20.\cite{sthesis} The structures for n=15, 16, 17, 18 and 20 were obtained by removing or adding atoms from the optimal 19-atom cluster structure and then letting them 
to optimize for all possible collinear spin configurations of the constituting atoms. In Fig. \ref{13to20}, we have shown a plot of binding energies of the optimal hcp and the optimal icosahedral structures of the Fe$_n$ and Co$_n$ clusters considering the clusters sizes in the range 15$\leq$n$\leq$20. It clearly indicates that the icosahedral growth pattern is more favorable for the small Fe$_n$ clusters and the hcp growth pattern for the small Co$_n$ clusters in the spin polarized calculations, in agreement with the trend observed for 19 atoms clusters discussed above. In Table II, we have also shown our estimated magnetic moments of the optimized hcp and icosahedral structures of Fe$_n$ and Co$_n$ clusters in this size ranges. Notice that our estimated magnetic moments for the optimized structures are in fair agreement with the recent result of Stern-Gerlach experiments for Fe-clusters\cite{FeExpt} and Co-clusters.\cite{CoExpt}

\begin{table}[h]

\begin{center}
\caption{\label{table2}Calculated magnetic moments of the optimized hcp and icosahedral structures of Fe$_{n}$ and Co$_{n}$ clusters (15$\leq$n$\leq$20) in spin-polarized calculations. For comparison, we have also listed the recent experimental values (Ref. 59 for Fe$_n$ clusters and Ref. 60 for Co$_n$ clusters) of magnetic moments in this size ranges.}
\vskip 0.2cm
\begin{tabular}{ccccc|ccccc}
\hline
\hline
Clusters  & & \multicolumn{2}{c} {M ($\mu_B$/atom)}&&Clusters && \multicolumn{2}{c}{M ($\mu_B$/atom)} \\
     
\cline{3-5}\cline{7-9}
          & &\multicolumn{2}{c}{Theory} & Expt.    &         & \multicolumn{2}{c}{Theory}   &   Expt. \\
\cline{3-4}\cline{7-8}
          & & hcp        & icosa        &          &         & hcp     & icosa   &    \\
\hline
Fe$_{15}$  &  &  3.07    &    3.20      & 2.72     &Co$_{15}$ & 2.07  &   1.93 & 2.38     \\
Fe$_{16}$  &  &  3.13    &    3.13      & 2.94     &Co$_{16}$ & 2.13  &   1.88 & 2.53     \\
Fe$_{17}$  &  &  3.18    &    3.06      & 2.86     &Co$_{17}$ & 2.06  &   2.06 & 2.24     \\
Fe$_{18}$  &  &  3.11    &    3.11      & 3.02     &Co$_{18}$ & 2.00  &   2.00 & 2.07     \\
Fe$_{19}$  &  &  3.05    &    3.05      & 2.92     &Co$_{19}$ & 2.05  &   1.95 & 2.21     \\
Fe$_{20}$  &  &  3.00    &    3.00      & 2.73     &Co$_{20}$ & 2.00  &   1.90 & 2.04     \\
\hline
\hline
\end{tabular}
\end{center}
\end{table}

In order to understand the optimal structures and the distortions in the structure that arises during optimization procedure, in Table III, we list the rms distortion of the bondlengths in the optimized geometries, which gives us 
the feel of the distortions that accompany optimization.

\begin{table}
\begin{center}
\caption{ Calculated root mean square (rms) deviations of bond lengths of Co$_{n}$ and Fe$_{n}$ clusters for n= 15 - 20, for both hcp and icosahedral symmetry
based optimized structures.}
\vskip 0.2cm
{\begin{tabular}{|ccc|ccc|}
\hline
\multicolumn{3}{|c|}{ Magnetic Co$_n$ clusters }& \multicolumn{3}{|c|} {Magnetic Fe$_n$ clusters} \\
\hline
Cluster & \multicolumn{2}{c|}{\it rms distortions} & Cluster & \multicolumn{2}{c|}{\it rms distortion} \\
size    &  hcp          &     icosa                  &size     &      hcp       &   icosa                 \\
\hline
Co$_{15}$&  0.036       &   0.073              &  Fe$_{15}$    &  0.100         &  0.075  \\
Co$_{16}$&  0.058       &   0.077              &  Fe$_{16}$    &  0.125         &  0.083  \\
Co$_{17}$&  0.054       &   0.110              &  Fe$_{17}$    &  0.129         &  0.076  \\
Co$_{18}$&  0.046       &   0.118              &  Fe$_{18}$    &  0.113         &  0.089  \\
Co$_{19}$&  0.043       &   0.073              &  Fe$_{19}$    &  0.121         &  0.096  \\
Co$_{20}$&  0.072       &   0.087              &  Fe$_{20}$    &  0.123         &  0.095  \\
\hline
\end{tabular} }
\end{center}
\end{table}

\begin{figure}
\begin{center}
\rotatebox{0}{\includegraphics[height=3.5cm,keepaspectratio]{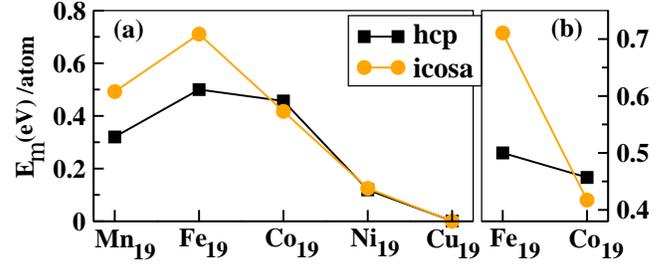}}
\caption{(Color online) (a) Magnetic energy (E$_m$), calculated as the energy difference between the spin polarized and non-spin polarized calculations of the optimized hcp (solid squares) and icosahedral (light solid dots) structures plotted for each X$_{19}$ cluster (X = Mn, Fe, Co, Ni, Cu). (b) Zoomed plot around the data points for the Fe$_{19}$ and Co$_{19}$ clusters.}
\label{mag_energy}
\end{center}
\end{figure}

The pertinent question, therefore, is what drives this phenomena ? To see the effect of magnetism, we first calculated the magnetic energy which is defined as the energy difference between the magnetic (spin-polarized) and nonmagnetic (non-spin-polarized) calculations for each of the hcp and icosahedral structures of X$_{19}$ clusters, estimated for their optimal structures in magnetic and nonmagnetic calculations. Fig. \ref{mag_energy}(a) shows the plot of magnetic energies of the X$_{19}$ clusters for the hcp and icosahedral symmetry based optimal structures. It is interesting to note that the magnetic energy of icosahedral structure is much higher than that of the hcp symmetry based structure for the Fe$_{19}$ clusters (and also for Mn$_{19}$ cluster, though we do not bring it into our discussion due to the assumption of collinearity in our calculation as mentioned before). On the other hand, it is of similar magnitudes for the hcp symmetry based structure of Co$_{19}$ cluster and its icosahedral counterpart, with hcp being somewhat higher. The magnetic energy difference between hcp and icosahedral structures is negligibly small in case of the Ni$_{19}$ and Cu$_{19}$ clusters. We have shown the zoomed plot again around the Fe$_{19}$ and Co$_{19}$ data points in Fig. \ref{mag_energy}(b) in order to see the effect of magnetic energy more closely, which shows opposite trend of magnetic energy gain between Fe$_{19}$ and that of Co$_{19}$ more clearly. We note that the difference of magnetic energy gains between the hcp and icosahedral structures in the case of the Co$_{19}$ cluster, is relatively small compared to that of the Fe$_{19}$ cluster. As the $d$-shell gets progressively filled up, starting from the half-filled situation with the highest atom-centered magnetic moment for Mn$_n$ cluster, the magnetic energy gain gets progressively weaker, thereby the role of magnetism being more important for Fe, compared to Co.

\begin{figure*}
\begin{center}
\rotatebox{0}{\includegraphics[height=12.5cm,keepaspectratio]{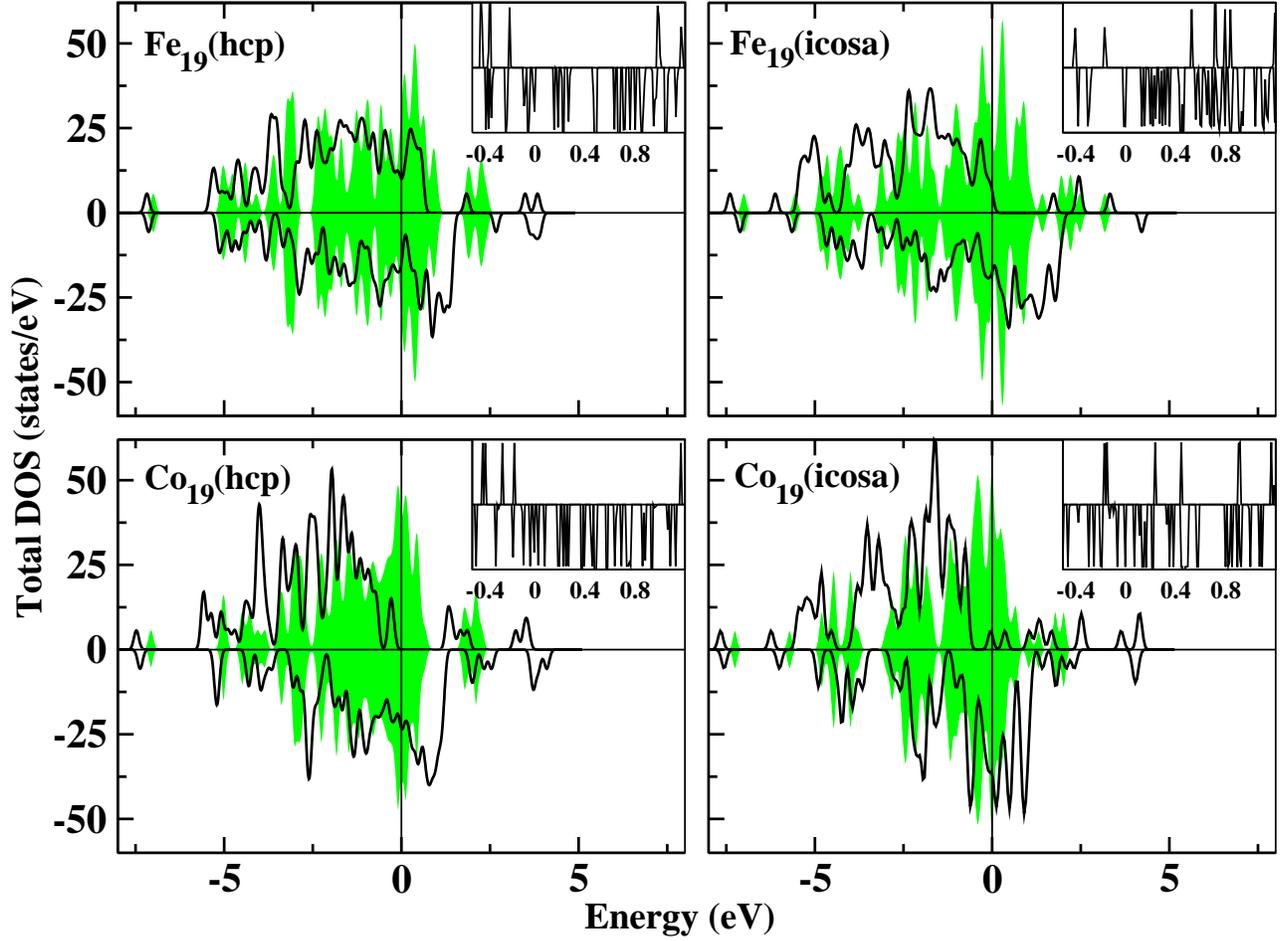}}
\caption{(Color online) Plot of DOS of the optimized hcp and icosahedral symmetry based structures of Fe$_{19}$ cluster (upper panels) and Co$_{19}$ cluster (lower panels) in spin-polarized as well as non-spin polarized calculations. The results for nonmagnetic calculations have been shown with shaded curves. The smearing width is fixed at 0.1 eV. The vertical line through zero is the Fermi energy for each system. The inset shows the DOS corresponding to the spin-polarized calculations, zoomed around the Fermi energy with a smearing of 0.001 eV.}
\label{dos}
\end{center}
\end{figure*}

In order to understand the gain in magnetic energy for the icosahedral structure of the Fe$_{19}$ cluster and for the hcp structure of the Co$_{19}$ cluster, we have studied the density of states (DOS) of the optimized hcp and icosahedral structures of the Fe$_{19}$ and Co$_{19}$ clusters for both  the magnetic and nonmagnetic calculations as shown in Fig. \ref{dos}. We note that compared to the nonmagnetic DOS, the gap in the majority spin channel is significantly enhanced in case of the icosahedral structure of Fe$_{19}$ and the hcp structure of Co$_{19}$, indicating their enhanced stability. On the other hand, for the optimal hcp structure of Fe$_{19}$ and for the optimal icosahedral structure of Co$_{19}$ cluster in case of spin-polarized calculation, there are finite amount of states around the Fermi energy, which reduces the stability of the system, compared to that of the corresponding icosahedral and hcp structures.

\begin{figure*}[h]
\begin{center}
\vskip 1.0cm
\rotatebox{0}{\includegraphics[height=5.5cm,keepaspectratio]{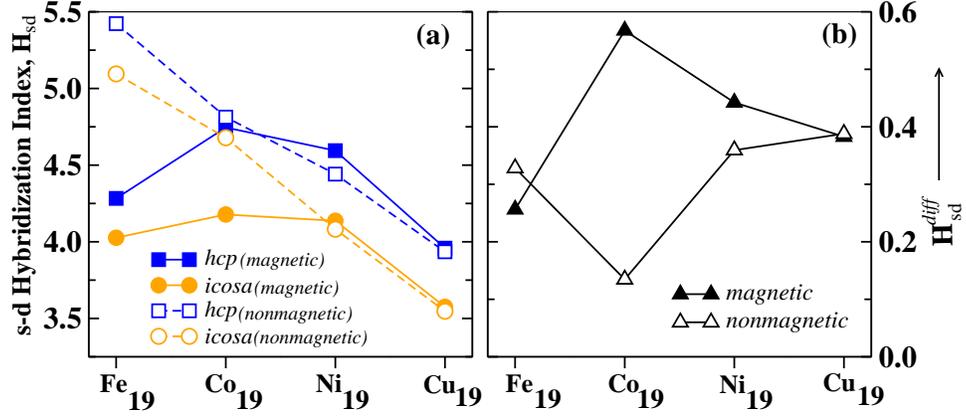}}
\caption{(Color online) Plot of (a) $s$-$d$ hybridization index for the optimized hcp and optimized icosahedral 19-atoms clusters of Fe, Co, Ni and Cu, both in the spin-polarized (magnetic) calculation and non-spin polarized (nonmagnetic) calculation (solid circles and squares represent the results for spin-polarized calculation and empty circles and squares correspond to the data points for non-spin polarized calculations), (b) the difference in $s$-$d$ hybridization (H$_{sd}$$^{diff}$=H$_{sd}$$^{hcp}$-H$_{sd}$$^{icosa}$) between the optimized hcp and icosahedral structures of 19-atom clusters both in the spin-polarized and non-spin polarized calculations.  }
\label{sd-diff}
\end{center}
\end{figure*}

We next study another relevant quantity which has been used previously to examine the relative stability between the various classes of isomers for the 3$d$ late transition metal clusters, namely the hybridization of the atomic 3$d$ and 4$s$ orbitals. The $s$-$d$ hybridization index as quantified by H$\ddot{a}$kkinen {\it et al}\cite{ref52} and later used by Chang {\it et al}\cite{ref53} as well as Wang {\it et al}\cite{ref54} for transition metal clusters, is defined for a 19 atoms cluster as
\begin{equation*}
H_{sd}=\sum\limits_{I=1}^{19}\sum\limits_{i=1}^{occ}w_{i,s}^{(I)}w_{i,d}^{(I)}
\end{equation*}
where $w_{i,s}^{I}$ ($w_{i,d}^{I}$) is the projection of $i$-th Khon-Sham orbital onto the $s$ ($d$) spherical harmonic centered at atom $I$, integrated over a sphere of specified radius. The spin index is implicit in the summation. Our calculated $s$-$d$ hybridization index for the optimized structures of both the symmetries
 for the Fe$_{19}$, Co$_{19}$ and also for the Ni$_{19}$, Cu$_{19}$ clusters have been plotted in Fig. \ref{sd-diff}. To see the effect of magnetism, we have studied the $s$-$d$ hybridization of the optimized structure of each cluster for both the magnetic and nonmagnetic phases.

It is seen that the optimized hcp structures have consistently higher values of H$_{sd}$ than those of the optimized icosahedral structures of  3$d$ late transition metal clusters X$_{19}$ for both spin-polarized and non-spin polarized calculations. In order to see distinctly the effect of magnetization on $s$-$d$ hybridization, we have plotted the difference of $s$-$d$ hybridization indexes between the optimized hcp and the optimized icosahedral structures for both the magnetic and nonmagnetic calculations in the right panel of Fig. \ref{sd-diff}. The positive (negative) value of this difference, H$_{sd}$$^{diff}$, indicates that the hcp (icosahedron) structure has higher $s$-$d$ hybridization. It is clearly seen that though the difference is positive for all the late transition metal clusters, it shows some variation across the series. The $s$-$d$ hybridization gain in favor of hexagonal structure is the maximum for the magnetic Co$_{19}$ cluster, showing a factor of about 6 times enhancement compared to nonmagnetic Co$_{19}$. Cu$_{19}$ cluster being essentially nonmagnetic, $s$-$d$ hybridization gain between the two structural symmetries remains same both in magnetic and nonmagnetic calculation of Cu$_{19}$. The $s$-$d$ hybridization gain remains similar for the magnetic Fe$_{19}$ and nonmagnetic Fe$_{19}$ (H$_{sd}$$^{diff}$ $\sim$0.3) and that for magnetic Ni$_{19}$ and nonmagnetic Ni$_{19}$ (H$_{sd}$$^{diff}$ $\sim$0.4). We therefore, conclude that the gain in $s$-$d$ hybridization stabilizes the hcp symmetry based structure over the icosahedral symmetry based structure for the Co$_{19}$ cluster. This is also helped in a way by the small but positive magnetic energy gain in favor of hcp phase of the Co$_{19}$ cluster. So the $s$-$d$ hybridization helped by magnetic energy gain stabilizes the hcp symmetry based structure in case of the Co$_{19}$ cluster. On the other hand, for Fe$_{19}$ cluster, the large magnetic energy gain in favor of the icosahedral symmetry, decides the final stability, thereby counteracting the hybridization energy gain in favor of hexagonal symmetry.
\section{Summery and Conclusions}
To summarize, we have investigated the relative stability of the 3$d$ late transition metal clusters specially of 19 atoms between  hcp and icosahedral symmetries. Among all the members, the Co$_{19}$ cluster prefers an unusual stabilization in hexagonal symmetry, while the rest show the preference of icosahedral symmetry. Our study nicely demonstrates that this curious result is driven by the interplay of the gain in magnetic energy {\it vis a vis} the gain in hybridization energy. For the Co$_{19}$ clusters, the hybridization energy gain helped by magnetic energy gain favors the stabilization of hexagonal symmetry, while for clusters like Fe$_{19}$, the large magnetic energy gain in icosahedral symmetry topples the $s$-$d$ hybridization gain in favor of hexagonal symmetry and stabilizes the icosahedral phase. We find the obtained trend to hold good as well for clusters having sizes between 15 and 20.

\acknowledgments
 T.S.D. and S. D. thank Department of Science and Technology, India for the support through
Advanced Materials Research Unit.

\end{document}